\documentclass[twocolumn,superscriptaddress,letter]{revtex4}%
\usepackage{amssymb}
\usepackage{color}
\usepackage{graphicx}
\usepackage{dcolumn}
\usepackage{bm}
\usepackage[header,title,page,titletoc]{appendix}
\usepackage{amsmath}
\usepackage{amsfonts}%
\providecommand{\U}[1]{\protect \rule{.1in}{.1in}}

\begin{document}
\title{Magnetic Orders of Correlated Topological Insulators at Finite Temperature}
\author{Ying-Xue Zhu}
\affiliation{Department of Physics, Beijing Normal University, Beijing, 100875, P. R. China}
\author{Jing He}
\affiliation{Department of Physics, Beijing Normal University, Beijing, 100875, P. R. China}
\author{Chun-Li Zang}
\affiliation{Department of Physics, Beijing Normal University, Beijing, 100875, P. R. China}
\author{Ying Liang}
\affiliation{Department of Physics, Beijing Normal University, Beijing, 100875, P. R. China}
\author{Su-Peng Kou}
\thanks{Corresponding author}
\email{spkou@bnu.edu.cn}
\affiliation{Department of Physics, Beijing Normal University, Beijing, 100875, P. R. China}

\begin{abstract}
In this paper, we study the magnetic orders of two dimensional correlated
topological insulators including the correlated Chern insulator and the
correlated $Z_{2}$ topological insulator at finite temperature. For the 2D
correlated Chern insulator, we found that thermal-fluctuation-induced magnetic
order appears in the intermediate interaction region of the correlated Chern
insulator. On the contrary, for the correlated $Z_{2}$ topological insulator
there doesn't exist the thermal-fluctuation-induced magnetic order. In the
end, we give an explanation on the difference.

PACS numbers:

\end{abstract}
\maketitle

In recent years, the physics community has witnessed a series of exciting
discoveries. Among them, topological insulator (TI) is quite impressive and
has become a rapidly-developing field. As the first example, the integer
quantum Hall (IQH) effect is a remarkable achievement in condensed matter
physics\cite{2,qhe}. To describe the IQH effect, the Chern number or so called
TKNN number, $C$, is introduced by integrating over the Brillouin zone (BZ) of
the Berry field strength \cite{thou}. So, this type of topological insulator
with IQH effect is called the Chern insulator. Recently, a new class of
topological insulator with time-reversal symmetry is discovered with the
quantized spin Hall effect\cite{kane,berg}. To label this class of TI, Kane
and Mele proposed a $Z_{2}$ topological invariant\cite{kane}. So, it is called
$Z_{2}$ topological insulator. For all these TIs, thermal fluctuations will
wash out their topological features. Therefore, to observe the topological
properties in a TI, low temperature is necessary condition. In addition,
people studied the correlated Chern insulator\cite{he1,he2,he3} and the
correlated $Z_{2}$ topological insulator\cite{ra,ho,xie,Zheng,liu,mar,zong} at
zero temperature by using different approaches and found they have much
difference properties.

On the other hand, in condensed matter physics, people have found different
types of (long range) orders (magnetic order, superconducting order, etc.). To
describe these ordered phases, different local order parameters are defined.
In general, thermal fluctuations will destroy the long range orders and drive
the system into a disordered phase. In Landau's theory, the phase transition
between an ordered phase and a disordered phase is always accompanied by
symmetry breaking.

In this paper, we will study the two dimensional (2D) correlated topological
insulators including the correlated Chern insulator and the correlated $Z_{2}$
topological insulator at finite temperature. For the 2D correlated Chern
insulator, the ground states can be a new type of topological quantum state -
topological spin-density-wave (TSDW) state\cite{he2}. An interesting issue is
\emph{the properties of TSDWs at finite temperature}. In particular, we found
that thermal-fluctuation-induced magnetic order appears in the intermediate
interaction region of the correlated Chern insulator. On the contrary, for the
correlated $Z_{2}$ topological insulator there doesn't exist TSDW and the
thermal-fluctuation-induced magnetic order.

\textit{The correlated Chern insulator}: Firstly we study the properties of
the 2D correlated Chern insulator, of which the Hamiltonian
is\cite{Haldane,he1,he2,he3}
\begin{equation}
H=H_{\mathrm{H}}+H^{\prime}+U\sum \limits_{i}\hat{n}_{i\uparrow}\hat
{n}_{i\downarrow}-\mu \sum \limits_{\left \langle i,\sigma \right \rangle }\hat
{c}_{i\sigma}^{\dagger}\hat{c}_{i\sigma}+h.c.
\end{equation}
where $H_{\mathrm{H}}$ is the Hamiltonian of the spinful Haldane model on a
honeycomb lattice which is given by
\begin{equation}
H_{\mathrm{H}}=-t\sum \limits_{\left \langle {i,j}\right \rangle ,\sigma}\left(
\hat{c}_{i\sigma}^{\dagger}\hat{c}_{j\sigma}+h.c.\right)  -t^{\prime}%
\sum \limits_{\left \langle \left \langle {i,j}\right \rangle \right \rangle
,\sigma}e^{i\phi_{ij}}\hat{c}_{i\sigma}^{\dagger}\hat{c}_{j\sigma}.
\end{equation}
Here $t$ and $t^{^{\prime}}$ are the nearest neighbor (NN) hopping and the
next-nearest neighbor (NNN) hopping, respectively. There exists a complex
phase $\phi_{ij}$ into the NNN hopping which is set to be the direction of the
positive phase clockwise $\left(  \left \vert \phi_{ij}\right \vert =\frac{\pi
}{2}\right)  $. $H^{\prime}$ denotes an on-site staggered energy which is
$H^{\prime}=\varepsilon \sum \limits_{i\in{A,}\sigma}\hat{c}_{i\sigma}^{\dagger
}\hat{c}_{i\sigma}-\varepsilon \sum \limits_{i\in{B,}\sigma}\hat{c}_{i\sigma
}^{\dagger}\hat{c}_{i\sigma}.$ $U$ is the on-site Coulomb repulsion strength.
$\left \langle {i,j}\right \rangle $ and $\left \langle \left \langle
{i,j}\right \rangle \right \rangle $ denote two sites of the NN and the NNN
links, respectively. $\hat{n}_{i\uparrow}$ and $\hat{n}_{i\downarrow}$ are the
number operators of electrons with up-spin and down-spin respectively. $\mu$
is the chemical potential and $\mu=U/2$ at half-filling for our concern in
this paper.

For free electrons, $U=0$, we can see that there exist energy gaps
$\Delta_{f1}$, $\Delta_{f2}$ near the two Dirac points $\mathbf{k}_{1}%
=-\frac{2\pi}{3}(1,$ $1/\sqrt{3})$ and $\mathbf{k}_{2}=\frac{2\pi}{3}(1,$
$1/\sqrt{3})$ as $\Delta_{f1}=\left \vert 2\varepsilon-6\sqrt{3}t^{\prime
}\right \vert $ and $\Delta_{f2}=2\varepsilon+6\sqrt{3}t^{\prime},$
respectively. There exist two phases separated by the phase boundary
$\Delta_{f1}=0$: the Chern insulator with Chern number $C=2$ and the normal
band insulator (NI) state. In the Chern insulator, due to the quantum
anomalous Hall (QAH) effect with a quantized (charge) Hall conductivity
$\sigma_{H}=2e^{2}/h,$ we denote the Chern insulator by "QAH".

\textit{Mean field approach}: With the increasing of the interaction, the
correlated Chern insulator is unstable against an antiferromagnetic (AF)
spin-density-wave (SDW) which is described by $\left \langle \hat{c}_{i,\sigma
}^{\dagger}\hat{c}_{i,\sigma}\right \rangle =\frac{1}{2}[1+(-1)^{i}\sigma M]$
where the local order parameter $M$ is the staggered magnetization. We set
$\sigma=+1$ for spin up and $\sigma=-1$ for spin down.

\begin{figure}[ptb]
\begin{center}
\includegraphics[width=4in]{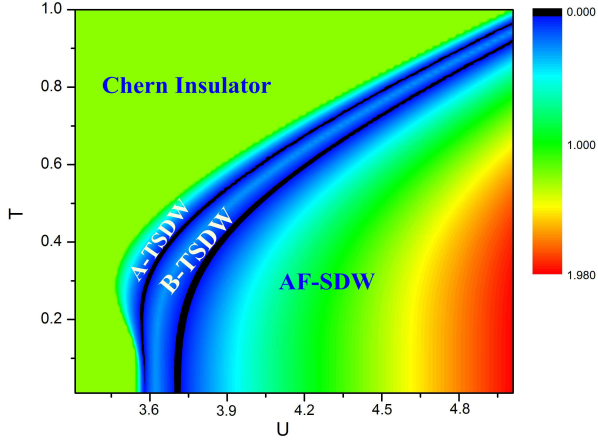}
\end{center}
\caption{(color online) The phase diagram of the correlated Chern insulator at
finite temperature for the case of $t^{\prime}/t=0.15$, $\varepsilon/t=0.15$.
There exist four phases: $C=2$ Chern insulator, A-TSDW, B-TSDW and trivial
AF-SDW. The color shows the energy gap of the electrons. }%
\end{figure}

Using the mean-field approach, we can obtain the self-consistent equation for
$M$ by minimizing the free energy at finite temperature in the reduced
Brillouin Zone (BZ):
\begin{align}
1  &  =\frac{1}{N_{s}M}\sum \limits_{\mathbf{k}}[\frac{\xi_{k}^{^{\prime}%
}+\varepsilon+\Delta_{M}}{2E_{1,k}}\tanh(\beta E_{1,k}/2)\nonumber \\
&  -\frac{\xi_{k}^{^{\prime}}+\varepsilon-\Delta_{M}}{2E_{2,k}}\tanh(\beta
E_{2,k}/2)] \label{self}%
\end{align}
where $N_{s}$ is the number of unit cells, $\beta=1/k_{\mathrm{B}}T$ and
$\Delta_{M}=UM/2$. Then the energy spectrums of electrons are $E_{1,k}%
=\sqrt{(\xi_{k}^{^{\prime}}+\varepsilon+\Delta_{M})^{2}+\left \vert
\xi_{\mathbf{k}}\right \vert ^{2}}$ and $E_{2,k}=\sqrt{(\xi_{k}^{^{\prime}%
}+\varepsilon-\Delta_{M})^{2}+\left \vert \xi_{\mathbf{k}}\right \vert ^{2}}$
where
\begin{align}
\xi_{k}  &  =t\sqrt{3+2\cos{(\sqrt{3}k_{y})}+4\cos{(3k_{x}/2)}\cos{(\sqrt
{3}k_{y}/2)}}\\
\xi_{k}^{^{\prime}}  &  =2t^{\prime}(\sin(\sqrt{3}k_{y})-4\cos(3k_{x}%
/2)\sin{(\sqrt{3}k_{y}/2)}).\nonumber
\end{align}

\textit{Phase diagram at finite temperature}: To determine the phase diagram
at finite temperature, there exist two types of phase transitions: one is the
"magnetic" phase transition that separates the magnetic order state with
$M\neq0$ and the nonmagnetic state with $M=0$ (solving the Eq.(\ref{self})),
the other one is the "topological" phase transition that is characterized by
the condition of zero fermion-energy gaps, $\Delta_{f}=\left \vert -6\sqrt
{3}t^{\prime}+2\varepsilon \pm2\Delta_{M}\right \vert =0$ (see the black lines
in Fig. 1). In Fig. 1, the colors show the energy gap of the electrons. After
determining the phase boundaries, we get the phase diagram at finite
temperature in Fig. 1 for the parameters of $t^{\prime}/t=0.15$,
$\varepsilon=0.15t.$

From Fig. 1, we can see that for the 2D correlated Chern insulator there exist
four phases: Chern insulator (QAH), A-TSDW (TSDW with Chern number $C=2$),
B-TSDW (TSDW with Chern number $C=1$) and trivial AF-SDW. The Chern insulator
exists in the weak-interaction region. With the increasing of $U/t$, the
system turns into A-TSDW state. After the electron's energy gap is closed at
one Dirac point, the system turns into B-TSDW state. With the interaction
strength further increasing, the electron's energy gap is closed at another
Dirac point and the system turns into a trivial AF-SDW state.

\begin{figure}[ptb]
\begin{center}
\includegraphics[width=3.5in]{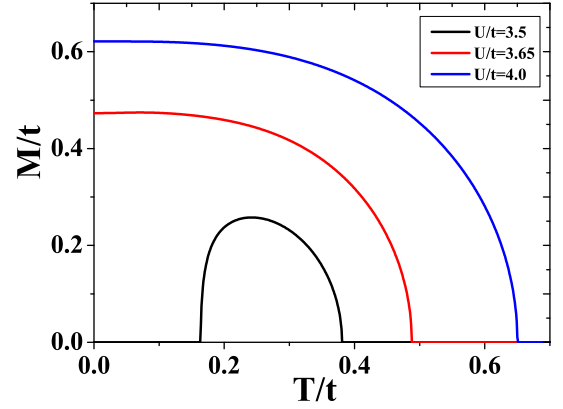}
\end{center}
\caption{(color online) The staggered magnetization $M$ via the temperature at
different interaction strength $U/t$ for the case $t^{\prime}/t=0.15$,
$\varepsilon=0.15t.$}%
\end{figure}

In the phase diagram, we find an interesting phenomenon: \emph{the
thermal-fluctuation-induced magnetic order}. From Fig. 1, there exists a
magnetic order at finite temperature for about $U/t=3.5$ during the
temperature $T/t=0.1\sim0.5$\cite{note}. In Fig. 2, we plot the staggered
magnetization $M$ with the increasing of temperature via the interaction
strengths $U$ for the case of $t^{\prime}/t=0.15$, $\varepsilon=0.15t$. From
Fig. 2, we can see that at $U/t=3.5$ (the black line), the staggered
magnetization $M$ is zero at low temperature. But at higher temperature, the
staggered magnetization $M$ becomes nonzero and has the maximum value at
$T/t\simeq0.25$ and then with the further increasing of temperature the
staggered magnetization $M$ becomes smaller and smaller down zero. Thus, this
magnetic order at finite temperature is assisted by the\ thermal fluctuations.

\begin{figure}[ptb]
\begin{center}
\includegraphics[width=3.5in]{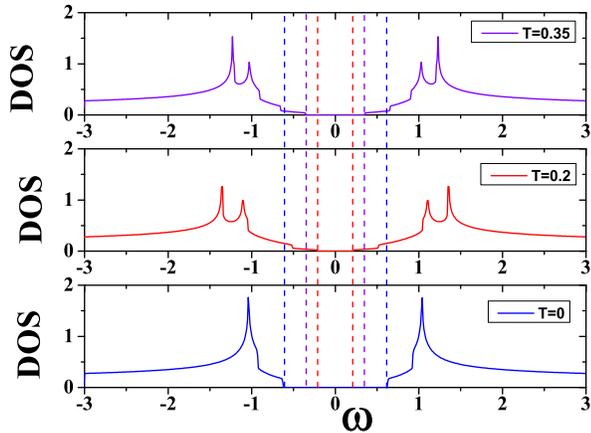}
\end{center}
\caption{(color online) The density of state (DOS) at different temperatures
for the case of $U/t=3.5,$ $t^{\prime}/t=0.15$, $\varepsilon=0.15.$}%
\end{figure}

Let's give a brief explanation on the existence of the
thermal-fluctuation-induced magnetic order. At zero temperature, the density
of state (DOS) inside the energy gap vanishes. At finite temperature, due to
the thermal fluctuations, the density of state (DOS) inside the energy gap
increases that may help the establish the magnetic order. In Fig. 3, we plot
the DOS at different temperatures for the case of $U/t=3.5,$ $t^{\prime
}/t=0.15$, $\varepsilon=0.15t.$ From Fig. 3, We can see that with the
increasing of the temperature, at first the magnetization increase, then the
gap of the system becomes smaller; when the temperature further increase, the
magnetization decreases, then the gap of the system becomes bigger again. This
result is confirmed by the results of the energy gap in Fig. 4(a) at
$U/t=3.5.$ In Fig. 4(b) and (c), we also give the energy gap for $U/t=3.65$
and $4.0$ for different temperatures for the case of $t^{\prime}/t=0.15$,
$\varepsilon=0.15t.$

\begin{figure}[ptb]
\begin{center}
\includegraphics[width=3.5in]{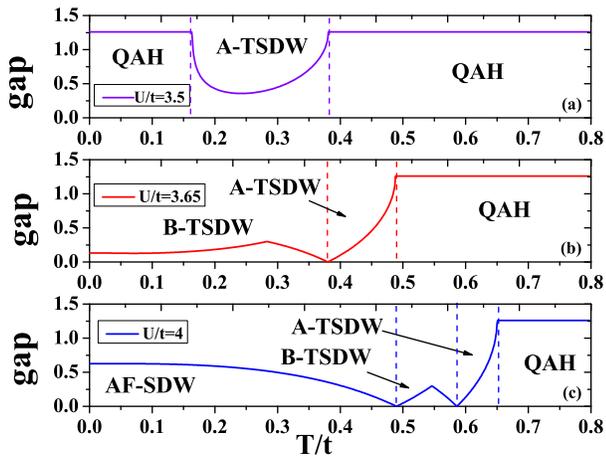}
\end{center}
\caption{(color online) The energy gap of electrons via the temperature for
the case of $t^{\prime}/t=0.15$, $\varepsilon=0.15t.$}%
\end{figure}

\textit{The "topological" phase transition at finite temperature}: A related
issue is about "topological" phase transition at finite temperature. To check
whether there exists true topological phase transition at finite temperature,
we calculate the Hall conductivity and the special heat.

We use the Kubo formula to derive the Hall conductivity\cite{qi}, $\sigma
_{H}=\lim_{\omega \rightarrow0}\frac{i}{\omega}Q_{xy}\left(  \omega
+i\delta \right)  $ where
\begin{align}
Q_{xy}\left(  i\nu_{m}\right)   &  =\frac{1}{N_{s}\beta}\sum_{k,n}%
\operatorname*{tr}[J_{x}\left(  k\right)  G\left(  k,i\left(  \omega_{n}%
+\nu_{m}\right)  \right) \nonumber \\
&  J_{y}\left(  k\right)  G\left(  k,i\omega_{n}\right)  ]
\end{align}
with the current operator $J_{x/y}\left(  k\right)  =\frac{\partial H\left(
k\right)  _{\sigma}}{\partial k_{x/y}}$ and $G\left(  k,i\omega_{n}\right)  $
is the Matsubara Green function, $\sigma$ is the spin index. In Fig. 5, we
show the Hall conductivity $\sigma_{H}$ via the interaction strength $U$ at
different temperatures for the case of $t^{\prime}/t=0.15$, $\varepsilon
=0.15t$. At zero temperature, we can use the Hall conductivity to characterize
the topological properties of the system. There exist three plateaus of the
Hall conductivity: $\sigma_{H}=2e^{2}/h$ in the Chern insulator (QAH) and
A-TSDW, $\sigma_{H}=e^{2}/h$ in B-TSDW, $\sigma_{H}=0$ in trivial AF-SDW. At
finite temperature, the situation changes. From Fig. 5, we can see that at
finite temperature, the Hall conductivity $\sigma_{H}$ smoothly changes with
the interaction and the temperature and the plateaus of the Hall conductivity
are smeared out. That means there is no true "topological" phase transition at
finite temperature.

In addition we calculate the special heat at finite temperature and also don't
find the true "topological" phase transition.

\begin{figure}[ptb]
\begin{center}
\includegraphics[width=3.5in]{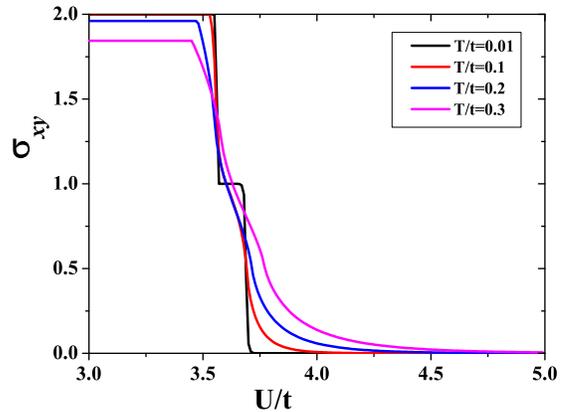}
\end{center}
\caption{(color online) The Hall conductivity $\sigma_{H}$ via the interaction
for the case of $t^{\prime}/t=0.15$, $\varepsilon=0.15t$ at different
temperatures. }%
\end{figure}

\textit{The correlated Kane-Mele model at finite temperature}: Next we study
the 2D correlated $Z_{2}$ topological insulator. Our start point is the
correlated Kane-Mele (KM) model which is described
by\cite{ra,ho,xie,Zheng,liu,mar,zong}
\begin{equation}
H=H_{\mathrm{KM}}+H^{\prime}+U\sum \limits_{i}\hat{n}_{i\uparrow}\hat
{n}_{i\downarrow}-\mu \sum \limits_{i}\hat{c}_{i}^{\dagger}\hat{c}_{i}%
\end{equation}
where $H_{\mathrm{KM}}$ is the Hamiltonian of the KM model which is given by%
\begin{equation}
H_{\mathrm{KM}}=-t\sum \limits_{\left \langle {i,j}\right \rangle }\left(
\hat{c}_{i}^{\dagger}\hat{c}_{j}+h.c.\right)  -t^{\prime}\sum
\limits_{\left \langle \left \langle {i,j}\right \rangle \right \rangle }%
e^{i\phi_{ij}}\hat{c}_{i}^{\dagger}\sigma_{z}\hat{c}_{j},
\end{equation}
and $H^{\prime}$ denotes an on-site staggered energy which is $H^{\prime
}=\varepsilon \sum \limits_{i\in{A,}\sigma}\hat{c}_{i\sigma}^{\dagger}\hat
{c}_{i\sigma}-\varepsilon \sum \limits_{i\in{B,}\sigma}\hat{c}_{i\sigma
}^{\dagger}\hat{c}_{i\sigma}.$ Here we set the on-site staggered energy
$\varepsilon$ to be $0.15t$.

Without the spin rotation symmetry, the staggered magnetic order is along
\textbf{XY}-plane. In this paper we take a staggered magnetic order along
\textbf{X}-direction as an example. Now we get the self-consistency equation
for $M$ by minimizing the free energy at temperature $T$ in the reduced
Brillouin zone as
\begin{align}
1  &  =\frac{1}{2N_{s}}\sum \limits_{\mathbf{k}}\frac{U}{2}{\{ \frac
{[1+\varepsilon(\Delta_{M}^{2}+\xi_{k}^{^{\prime}2})^{-\frac{1}{2}}]}%
{-E_{1,k}}\tanh{(}\frac{{\beta}E_{1,k}}{2}{)}}\nonumber \\
&  {+\frac{[1-\varepsilon(\Delta_{M}^{2}+\xi_{k}^{^{\prime}2})^{-\frac{1}{2}%
}]}{-E_{2,k}}\tanh{(}\frac{{\beta}E_{2,k}}{2}{)\}}} \label{self2}%
\end{align}
where $E_{1,k}=-\sqrt{(\sqrt{\Delta_{M}^{2}+\xi_{k}^{^{\prime}2}}%
+\varepsilon)^{2}+|\xi_{k}|^{2}},$ $E_{2,k}=-\sqrt{(\sqrt{\Delta_{M}^{2}%
+\xi_{k}^{^{\prime}2}}-\varepsilon)^{2}+|\xi_{k}|^{2}}$ and $\Delta_{M}=UM/2.$

\begin{figure}[ptb]
\begin{center}
\includegraphics[width=4in]{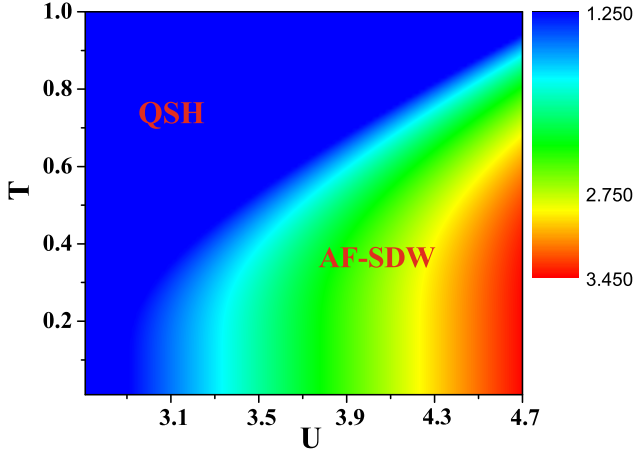}
\end{center}
\caption{(color online) The phase diagram of the correlated Kane-Mele model at
finite temperature for the case of $t^{\prime}/t=0.15$, $\varepsilon/t=0.15$.
There exist two phases: $Z_{2}$ topological insulator with quantized spin Hall
(QSH) effect with $M=0$ and trivial AF-SDW with $M\neq0$. The color denotes
the energy gap of electrons.}%
\end{figure}

In the phase diagram of correlated KM model, there only exists one phase
transition: the magnetic phase transition between a magnetic order state with
$M\neq0$ and a nonmagnetic state with $M=0$ (solving the Eq.(\ref{self2})). In
Fig. 6, we plot the phase diagram for the case of $t^{\prime}/t=0.15$,
$\varepsilon=0.15t.$ From the phase diagram, we can see that there exist two
phases: the $Z_{2}$ topological insulator with $M=0$ and trivial AF-SDW state
with $M\neq0$. In the $Z_{2}$ topological insulator, due to the nonzero
$Z_{2}$ topological invariant, there exists the quantum spin Hall (QSH)
effect. In this paper we denote the $Z_{2}$ topological insulator by "QSH".
From Fig. 6, we can see that the energy gap will be never closed. And there
doesn't exist the thermal-fluctuation-induced magnetic order.

\textit{Discussion and conclusions}: Finally, we discuss why the properties of
the correlated Chern insulator are much difference from those of the
correlated $Z_{2}$ topological insulator.

For the SDW order in the correlated Chern insulator, the effective Hamiltonian
becomes%
\begin{equation}
H_{0}=v_{F}\mathbf{p}\cdot \boldsymbol{\alpha}+\left(  \mathbf{m}%
_{T}+\mathbf{m}_{M}+\varepsilon \right)  \beta \label{modified-Dirac}%
\end{equation}
where $p_{i}=-i\hbar \nabla_{i}$ is the momentum operator ($i\in \{x,y\}$),
$p^{2}=p_{x}^{2}+p_{y}^{2}$, $v_{F}$ is the Fermi-velocity. There exist two
mass matrices: the mass matrix of the parent topological insulator
$\mathbf{m}_{T}=3\sqrt{3}t^{\prime}\eta_{3}\otimes \mathbf{I}_{2}$ and the mass
matrix from the SDW order $\mathbf{m}_{M}=\Delta_{M}\left(  \mathbf{I}%
_{2}\otimes \sigma_{3}\right)  $, respectively. Here $\mathbf{\tau},$
$\mathbf{\sigma}$\ and $\mathbf{\eta}$\ are Pauli matrices that denote the
indices of the sublattice, spin, node, respectively. $\mathbf{I}_{2}$ is the
$2\times2$ unit matrix, and $\otimes$ represents the Kronecker product. The
Dirac matrices can be expressed as a set of $4\times4$ matrices $\alpha
_{i}=\tau_{i}\otimes \mathbf{I}_{2},\  \beta=\tau_{z}\otimes \mathbf{I}_{2}.$
Here we set the on-site staggered energy $\varepsilon$ to be $0.15t$. Due to
$\left[  \mathbf{m}_{T},\text{ }\mathbf{m}_{M}\right]  =0$, we have the energy
gap of the electrons to be $\Delta_{f}=2\left \vert \pm \mathbf{m}_{T}%
\pm \mathbf{m}_{M}-2\varepsilon \right \vert =\left \vert \pm6\sqrt{3}t^{\prime
}\pm2\Delta_{M}-2\varepsilon \right \vert $. Thus, the energy gap from the
magnetic order $\mathbf{m}_{M}$ competes with that from the parent topological
insulator $\mathbf{m}_{T}$. Therefore, we call this system \emph{the
mass-gap-competition}. When we have a small staggered magnetization (or a
small $\Delta_{M}$), the energy gap of the electrons shrinks. With increasing
the staggered magnetization, the energy gap will eventually close at the
critical point $\Delta_{f}=0$. At finite temperature, the thermal fluctuations
will excite the quasiparticle and may also smear out the energy gap. Thus, due
to the suppression of the energy gap of the parent topological insulator, the
magnetic order may be assisted by thermal fluctuations.

For the SDW order in the correlated KM model, the effective Hamiltonian
becomes
\begin{equation}
H_{0}=v_{F}\mathbf{p}\cdot \boldsymbol{\alpha}+\left(  \mathbf{m}%
_{T}+\mathbf{m}_{M}+\varepsilon \right)  \beta
\end{equation}
where the mass term of the parent topological insulator is $\mathbf{m}%
_{T}=-3\sqrt{3}t^{\prime}\eta_{3}\otimes \sigma_{3}$ and the mass term from the
SDW order is $\mathbf{m}_{M}=\Delta_{M}\left(  \mathbf{I}_{2}\otimes \sigma
_{1}\right)  $, respectively. Due to $\{ \mathbf{m}_{T},$ $\mathbf{m}_{M}\}=0$,
we have the energy gap of the electrons to be $\Delta_{f}=\left \vert \pm
2\sqrt{\left \vert \mathbf{m}_{T}\right \vert ^{2}+\left \vert \mathbf{m}%
_{M}\right \vert ^{2}}-2\varepsilon \right \vert =\left \vert \pm \sqrt{(6\sqrt
{3}t^{\prime})^{2}+(\Delta_{M})^{2}}-2\varepsilon \right \vert $. From the
$Z_{2}$ topological insulator state of the correlated KM model ($(6\sqrt
{3}t^{\prime})-2\varepsilon>0$)\cite{kane}, when there exists the magnetic
order along \textbf{XY} plane, the energy gap of the electrons will definitely
increase. Therefore, the energy gap will never close in the magnetic order.
Therefore, we call this system \emph{the mass-gap-coexistence}. The
suppression of the energy gap of the parent topological insulator will not
help the formation of the magnetic order. Thus, there is no
thermal-fluctuation-induced magnetic order in the $Z_{2}$ topological
insulator of the correlated KM model.

At the end of the paper, we generalize the results in the correlated
topological insulators to a long range order developed in an insulator. For
the long range orders developed in an insulator, there are two different
cases: case I, the mass-gap-competition; case II, the mass-gap-coexistence.
For case I, the mass gap induced by a long range order $\mathbf{m}_{O}$
competes with the mass gap in the parent insulator $\mathbf{m}_{T}$. Now we
have $\left[  \mathbf{m}_{T},\text{ }\mathbf{m}_{M}\right]  =0$ and the energy
gap of the system is $\Delta_{f}=2\left \vert \mathbf{m}_{T}\pm \mathbf{m}%
_{M}\right \vert $. For this case, there may exist the
thermal-fluctuation-induced order. For case II, the mass gap induced by a long
range order $\mathbf{m}_{O}$ coexists with the mass gap in the parent
insulator $\mathbf{m}_{T}$. Now we have $\{ \mathbf{m}_{T},$ $\mathbf{m}%
_{M}\}=0$ and the energy gap of the system is $\Delta_{f}=2\sqrt{\left \vert
\mathbf{m}_{T}\right \vert ^{2}+\left \vert \mathbf{m}_{M}\right \vert ^{2}}$.
For this case, there doesn't exist the thermal-fluctuation-induced order.

\begin{acknowledgments}
This work is supported by National Basic Research Program of China (973
Program) under the grant No. 2011CB921803, 2012CB921704, 2011cba00102, NFSC
Grant No.11174035.
\end{acknowledgments}


\begin{thebibliography}{99}
\bibitem {note}For the 2D correlated topological insulator with spin rotation
symmetry, there is no true long range SDW order at finite temperature. Here,
A-TSDW, B-TSDW, trivial AF-SDW orders at finite tempearure are all short range
magnetic orders. If we consider a quasi-2D system with weakly interlayer
coupling, the short range magnetic order may turn into true long range order.
\end{thebibliography}
\end{document}